\documentstyle[prl,aps,multicol]{revtex}
\input{epsf}
\begin{document}
\draft
\title{Quantum Zeno Effect in Quantum Chaotic Systems}

\author{Sang Wook Kim, Young-Tak Chough, and Kyungwon An}

\address{Center for Macroscopic Quantum-Field Lasers and
Department of Physics, \\ Korea Advanced Institute of Science $\&$
Technology, Taejon 305-701}

\date{29 September 2000}

\maketitle

\begin{abstract}

We analyzed the effect of frequent measurements on the quantum systems that are
chaotic in the classical limit. It is shown that the kicked rotator, a well-known example 
of quantum chaos,  is too special to be used as a testing ground for the effects of 
the repeated measurements. The abrupt change of state vectors by the delta-kick 
singular interruptions causes a quantum anti-Zeno effect.  However, in more realistic 
systems with interaction potentials of continuous time dependence the quantum Zeno 
effect prevails.

\end{abstract}

\pacs{PACS number(s): 05.45.Mt, 03.65.Bz}

\begin{multicols}{2}

\narrowtext

Quantum mechanical behavior of the systems that are chaotic in the
classical limit, so called quantum chaos, has recently been the subject of
considerable interest \cite{Reichl92}.
Now it is generally accepted that
classical-like chaos is absent in quantum mechanics.
We can consider dividing the whole physical problem of quantum
dynamics into two qualitatively different parts: unitary time evolution
of the wave function $\psi(t)$ described by Schr\"{o}dinger's equation 
and the collapse of the wave function caused by quantum measurements. 
The first part has been mostly investigated in the study of quantum chaos so far
while the second part still remains controversial. The absence of classical-like
chaos is agreed only in the first part. It has been stressed that, particularly
by Lamb, that the clarification of quantum chaos should be based on 
the thorough understanding of the concept of measurements in quantum 
mechanics \cite{Lamb85}.

One of the most paradoxical results in the measurement problem is the
quantum Zeno effect (or quantum watched pot) \cite{Misra77}, which is
the inhibition of the time evolution of a quantum system, 
from one eigenstate of an observable into a superposition of eigenstates, by 
frequently repeated measurements . It can occur when measurement are
repeated so rapidly that the time between any two successive measurements 
is much shorter than the natural life time of the state. In his 1930 book 
\cite{Dirac}, Dirac already asserted that an observation
of an observable always results in one of the eigenvalues
of that observable, and that two measurements of the same
observable made in rapid succession would give the same results.
Itano {\em et al.} experimentally found that Rabi oscillations
were diminished by frequent measurements of the survival of the
initial eigenstate, which was considered as the first demonstration of the
quantum Zeno effect \cite{Itano90}. However, it is arguable whether this
confirms the quantum Zeno effect completely 
\cite{Ballentine91,Frerichs91,Pascazio94}.
Recently Kofman and Kurizki \cite{Kofman00} found that the modification 
of a decay process is determined by the energy spread incurred
by the measurements and by the distributions of the states to which
the decaying state is coupled. They concluded, whereas the
inhibitory quantum Zeno effect may be feasible in a limited class
of systems, the opposite effect, accelerated decay or quantum
anti-Zeno effect, appears to be much more ubiquitous.

One major modification that quantum mechanics introduces
to the classical picture of the deterministic chaos is the suppression
of chaotic diffusion, a phenomenon usually referred to as
dynamical localization (DL). This phenomenon, first discovered by Casati
{\em et al.} \cite{Casati79} in their investigation of the kicked rotator, 
can be understood as a dynamical version of Anderson
localization in solids \cite{Fishman82}. Just like the other quantum interference 
effects, DL is very sensitive to any incoherent perturbation. 
In the quantum kicked rotator, even when the corresponding classical
dynamics is regular, a diffusive behavior is obtained if a 
measurement is performed after each kick \cite{Facchi99}.
This phenomena was interpreted 
as an example of the anti-Zeno effect \cite{Kaulakys97}. However,
in this paper we show, although the kicked rotor is a well-known example of the 
quantum chaos, it is too special to be used as a testing ground for the effect of 
the repeated measurements. The anti-Zeno effect in the kicked
rotator is rather a trivial consequence of singular interruptions by the delta kicks. 
In more general classes of the quantum systems that are chaotic in the classical limit,
the quantum Zeno effect prevails, which can be also confirmed by using the results of 
Kofman and Kurizki \cite{Kofman00}.

\begin{figure} 
\centerline{\epsfxsize=8cm\epsffile{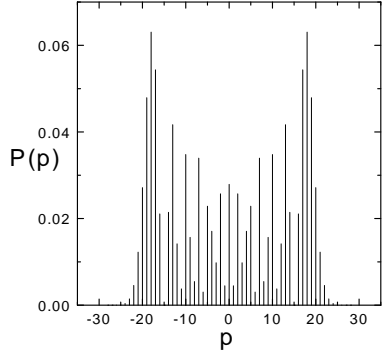}}
\caption{Probability distribution of the state vector
in the momentum space after a single kick in the kicked 
rotator with the initial state $|n=0 \rangle$ for $\lambda = 20$ 
and $T=1/4$. It is noted that the survival probability, 
$P(1) (=|c_0(1)|^2)$, is less than 3\%.}
\end{figure}

Let us begin by analyzing the kicked rotator described by the following Hamiltonian
\begin{equation}
\label{kicked rotator} H = \frac{p^2}{2} + \lambda \cos x
\sum_{k=-\infty}^{\infty} \delta (t-kT) 
,\end{equation} 
where $\lambda$ and $T$ is the kick strength and the time interval
between successive delta kicks, respectively. 
Using the complete set of eigenstates
$|n\rangle$, in which $\langle x|n\rangle = 
\exp(inx)/\sqrt{2\pi}$, the state just after the $(k+1)$th kick, $\psi(k+1) = \sum 
c_n(k+1)|n\rangle$, can be described by unitary transformation of the state just after 
the $k$th kick, $\psi(k) = \sum c_n(k)|n\rangle$:
\begin{equation}
c_n (k+1) = \sum_m U_{nm} c_m (k) 
,\end{equation} 
where $U_{nm} = (-i)^{n-m} J_{n-m}(\lambda/\hbar)
\exp(-i\hbar m^2 T/2)$ with $J_n$, the $n$th order Bessel function.
Figure 1 shows the probability distribution
of the state after a single kick for the initial condition 
$|n=0 \rangle$ with $\lambda = 20$
and $T=1/4$. Since during the time interval between successive kicks the momentum 
distribution is not changed, it is meaningless to perform the measurements more 
frequently than at the rate of $1/T$.

The quantum Zeno effect is represented by the relation,  $ P(t) \rightarrow 1$ as 
$\tau \rightarrow 0$, where $\tau$ and $P(t)$ represent the time interval between
successive measurements and the survival probability of the initial state, respectively. 
Note that the Zeno effect concerns itself with the short time behavior of the quantum evolution.  
In the kicked rotator the Zeno effect cannot take place by the following reasons.  
First, since the momentum distribution does not change during the time interval between 
successive kicks, any measurement done during that time interval, will trivially yield 
an identical result.
Hence, for any meaningful discussion of the Zeno effect, we should consider $\tau$ 
larger than the period $T$.  Once we have this constraint, the fact that $P(t)$ changes abruptly 
even after a single kick leads us to conclude that any physically meaningful Zeno effect 
cannot occur in this system.
The quantum localization due to the quantum interference is destroyed in the 
kicked rotator by the decoherence induced by the measurements.
This observation is interesting but does not prove the existence of the anti-Zeno 
effect in a broad class of the quantum systems that are chaotic in the classical limit.

\begin{figure} 
\centerline{\epsfxsize=8cm\epsffile{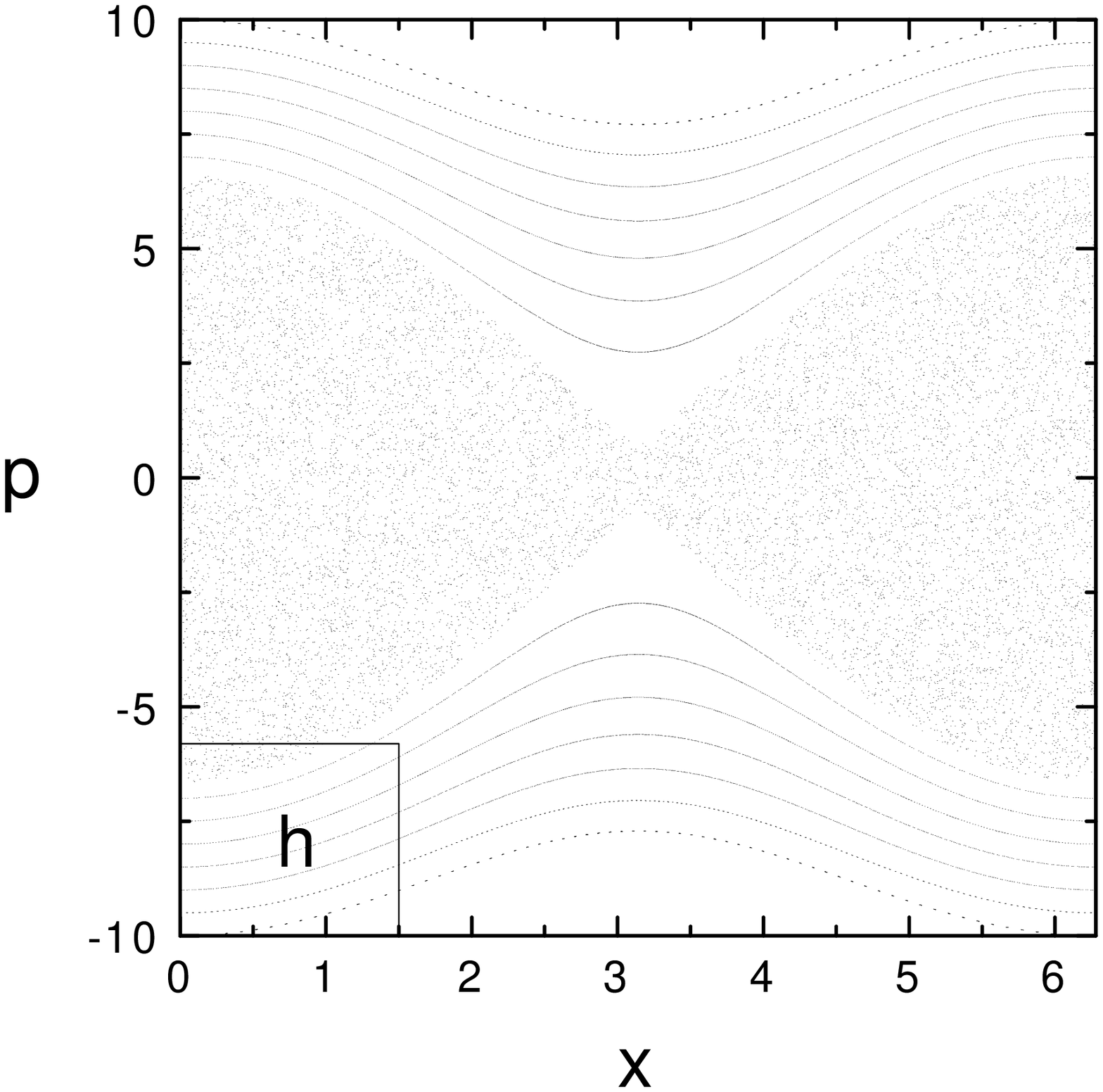}}
\caption{Partial Poincar\'{e} surface of section 
showing the phase space dynamics described by the Hamiltonian of 
Eq. (\ref{new rotator}) for $\lambda = 10$ and $T=2\pi$. 
The box in the left corner illustrates the size of Planck 
constant $h$.}
\end{figure}

In order to analyze the effect of repeated measurements in the absence of the abrupt 
changes of  the wave function as in  the kicked rotator, we consider a new Hamiltonian 
with continuous time dependence:
\begin{equation}
\label{new rotator}
H = \frac{p^2}{2} + \lambda \cos x \cos t
.\end{equation}
It is noted that the delta kicks in Eq.~(1) can be expanded as follows,
\begin{equation}
\label{delta}
\sum_{k=-\infty}^{\infty} \delta (t-kT) = \frac{2}{T}
\sum_{m=1}^{\infty} \cos(\frac{2\pi m t}{T}) + \frac{1}{T}
,\end{equation}
so that Eq.~(\ref{new rotator}) corresponds to the rotator driven by
a single frequency mode among the infinite numbers of modes of the kicked rotator.
Figure 2 shows a Poincare surface of section of Eq.~(\ref{new rotator})
with $\lambda = 10$, where a confined chaotic region is seen near
the origin. While the diffusion of the classical kicked rotator does not stop 
due to the infinite number of frequency components, the classical as well as 
the quantum diffusion in Eq.~(\ref{new rotator}) become saturated.
In Fig.~3 we show that both classical and quantum evolution of
$\Delta p^2 (t)$ have the same average level except for the quantum
fluctuations, so that both the quantum and classical results in Fig.~3 
correspond to a delocalized regime. In our discussion it is not necessary 
that the quantum diffusion should be suppressed below the classical one, 
the so-called localized regime.

\begin{figure} 
\centerline{\epsfxsize=8cm\epsffile{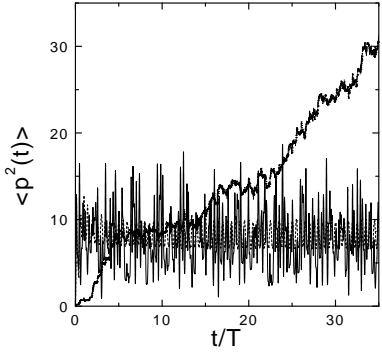}}
\caption{The evolution of momentum width $\Delta
p^2 (t)$ as a function of time in a unit of the period of driving
field in the system described by the Hamiltonian of Eq. (\ref{new rotator}). 
The solid and the dotted curves represent the pure 
quantum evolution and the measured quantum evolution with
$\tau = T/1000 $, respectively. The dashed curve 
corresponds to the classical evolution.}
\end{figure}

Consider now the dynamics of a system subjected to measurements
at time $n\tau$ ($n$ is an integer). The wave function can be expanded
into a linear superposition of complete bases of an observable, say,
$A$, for example
\begin{equation}
\label{eigen}
\psi(t) = \sum_n c_n (t) u_n, ~~~ A u_n = a_n u_n \;. 
\end{equation}
where $c_n (t) = |c_n (t)| e^{i\phi (t)}$ with $\phi(t)$ a phase factor.
A measurement of the system's
state at $t=n\tau$ projects the system onto one of the eigenstates
of the measured observable $A$ with a probability $|c_n(t)|^2$.
After  the measurement we know the probabilities, but we lose
information on the phase of the amplitudes $\phi(t)$,
so that the phase after the measurement can be regarded being random. 
This is nothing but the simple measurement postulate, so-called von 
Neumann's formulation of the wave function collapse.

\begin{figure} 
\centerline{\epsfxsize=8cm\epsffile{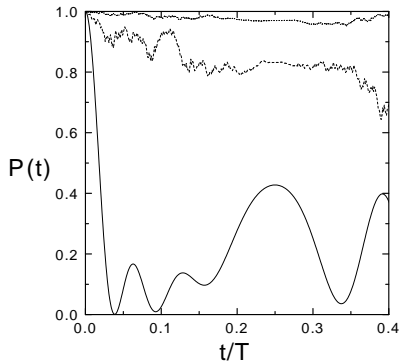}}
\caption{The survival probability for the pure
quantum evolution (the solid curve) and the measured quantum
with $\tau=T/1000$ (the dashed curve) and $\tau=T/5000$
(the dotted curve). This clearly shows quantum Zeno effect.}
\end{figure}

By using this concept we can find the quantum evolution of the system 
of Eq.~(\ref{new rotator}) under the repeated measurements. 
The system is described by a unitary transformation given by 
Schr\"{o}dinger's equation in the momentum representation.
Upon measuring the momentum we randomize the phases of
all the amplitudes, which can be also expressed by a unitary
transformation of a form of a diagonal matrix with random complex 
numbers of unity amplitude. Even though this diagonal matrix is unitary
in the mathematical expression, it can be considered irreversible in time 
in the physical sense since the phase randomization corresponds to
a stochastic process. In practice, for reducing the fluctuations
induced by the phase randomization, we averaged over 100
data sets for each result. Figure 4 shows that the autocorrelation
of the initial wave function (or survival probability),
$|\langle\psi (0)|\psi (t)\rangle|^2$,
without measurement (the solid curve) decreases faster than that 
under measurement (the dashed and the dotted curves). In addition,
the shorter is the time interval of the measurement, the slower is the 
decay of the autocorrelation. This is exactly what the quantum Zeno effect is.

Figure 3 presents $\Delta p^2 (t)$ for the quantum evolution both
with and without measurement, where the long time behavior is totally 
different from the short time one. The decoherence induced by the
measurements destroys quantum (and classical) localization.
It should be emphasized that this delocalization (or diffusion)
does not show anti-Zeno effect since Zeno effect manifests itself
only in the short time limt.
It is also noted that the measured quantum dynamics is not equivalent
to the classical dynamics regardless of the frequency of measurements, which
has been already known in several previous works
\cite{Facchi99,Dittrich90}.

To gain more physical insight let us examine the present problem using 
the analysis tool similar to that of Ref. \cite{Kofman00}.
We write $\hat{H} = \hat{H}_0  + \hat{V}g(t)$,
so that $\hat{H}_0 | \alpha \rangle = \hbar \omega_\alpha
| \alpha \rangle$ and $\hat{H}_0 | j \rangle =
\hbar \omega_j | j \rangle$, where $| \alpha \rangle$ is
an initial (decaying) state and ${| j \rangle}$'s are orthogonal to
$| \alpha \rangle$. The wavefunction can be written by
\begin{equation}
\label{wave function}
\psi(t) = \alpha(t) e^{-i\omega_\alpha t} | \alpha \rangle
+ \sum_{j} \beta_j e^{-i\omega_j t} | j \rangle
.\end{equation}
One can then obtain the following equations from Schr\"{o}dinger's 
equation 
\begin{eqnarray}
\label{eq. of motion 1}
\dot{\alpha} & = & -\frac{i}{\hbar}\sum_j 
\langle \alpha|\hat{V}|j \rangle
e^{i(\omega_\alpha - \omega_j )t} g(t) \beta_j \\
\label{eq. of motion 2}
\dot{\beta}_j & = & -\frac{i}{\hbar}
\langle j|\hat{V}|\alpha \rangle
e^{-i(\omega_\alpha - \omega_j )t} g(t) \alpha
.\end{eqnarray}
Integrating (\ref{eq. of motion 2}) and inserting it into
(\ref{eq. of motion 1}), we obtain the equation
\begin{equation}
\label{alpha dot}
\dot{\alpha} = - \int^{t}_{0} dt' e^{i\omega_\alpha (t-t')}
\Phi (t-t') g(t) g(t')\alpha(t')
,\end{equation}
where
\begin{equation}
\label{Phi(t)}
\Phi (t) = \hbar^{-2}\sum_{j}
| \langle \alpha|\hat{V}|j\rangle |^2
e^{-i\omega_j t}
.\end{equation}
Since we need to obtain only the short time behavior, for which $\alpha(t)
\approx \alpha(0) = 1$, we can set $\alpha(t') = 1$ in the right
side of Eq.~(\ref{alpha dot}) to obtain
\begin{equation}
\label{heart}
\alpha \simeq 1 - \int_0^t dx e^{i\omega_\alpha x}\Phi(x) M(t,x),
\end{equation}
where
\begin{equation}
M(t,x)= \int_x^t dt' g(t')g(t'-x)
.\end{equation}
If measurements are performed at sufficiently small interval $\tau$, 
we can obtain the survival probability, $P(t)$, using Eq.~(\ref{heart})
\begin{equation}
P(t=n\tau) \simeq |\alpha(\tau)|^{2n} \approx e^{-R(\tau)t}
,\end{equation}
where
\begin{equation}
\label{R(t)}
R = \frac{2}{\tau}Re\int_0^\infty dt e^{i\omega_\alpha t}
\Phi(t) M(\tau,t) \theta(\tau-t)
.\end{equation}
The function $\theta(x)$ is 1 for $x \geq 0$ and 0 for $x<0$. Using the 
convolution theorem of Fourier transform, Eq.~(\ref{R(t)}) can be rewritten by
\begin{equation}
\label{R(w)}
R=2\pi \int_0^\infty d \omega G(\omega) F(\omega)
,\end{equation}
where
\begin{equation}
\label{G(w)}
G(\omega) =   Four \{ \Phi(t) \}
= \hbar^{-2}\sum 
| \langle \alpha|\hat{V}|j \rangle |^2
\delta (\omega - \omega_j)
\end{equation}
and
\begin{equation}
\label{F(w)}
F(\omega) =  Four \left\{ \frac{1}{\tau} M(\tau,t)
e^{i\omega_\alpha t} \theta(\tau-t) \right\}
.\end{equation}
$Four(f)$ means Fourier transform of the function $f$.
While the function $G(\omega)$ in Eq.~(\ref{G(w)}) is the spectral
density of the states depending on the Hamiltonian,
especially the coupling, $\hat{V}$, the form factor $F(\omega)$
in Eq.~(\ref{F(w)}) shows the broadening of the eigenvalue $\omega_\alpha$
incurred by the frequent measurements. If the time independent
potential is considered, simply $F_0 (\omega) = 
(\tau/2\pi) {\rm sinc}^2 ((\omega-\omega_\alpha)\tau/2)$, 
in which the broadening rate is approximately $1/\tau$, 
consistent with the uncertainty principle.

In the case of a time dependent potential like the one described by 
Eq.~(\ref{new rotator}), only $F(\omega)$ is modified from $F_0 (\omega)$. 
Even though this is not just simple sinc function, the broadening of 
$\omega_\alpha$ is of the order of $1/\tau$ or larger. As the time interval of 
measurement, $\tau$, decreases, the broadening of $F(\omega)$ 
becomes larger, so that for a given $\omega$ the value of $F$ also decreases.
Consequently we can roughly write $F \sim \tau$ (or monotonically 
increasing function of $\tau$). 
Since $\langle 0 | \cos x | j \rangle = (\delta_{1j} +
\delta_{-1j})/2$ in (\ref{new rotator}),
we obtain the following simple equation
\begin{equation}
\label{G for new}
G(\omega) = \lambda^2\delta \left(\omega-\frac{1}{2} \right)
.\end{equation}
From these it can be easily shown that $R \sim \lambda^2 \tau$, 
which means that in the limit of $\tau \rightarrow 0$, 
the decay rate goes to zero. This corresponds to the Zeno effect. 
If the delta kicks in Eq.~(\ref{kicked rotator}) 
are replaced by pulses with a gaussian shape with finite time duration of
interaction, which also means more harmonics are included in 
Eq.~(\ref{new rotator}), the above discussion is still applicable, and thus we  find
the decay rate $R$ still proportional to $\tau$ with a different numerical constant.
Therefore, for a realistic kicked rotator we reach the same conclusion, the Zeno effect.

In summary, we have investigated the effect of frequent measurements on quantum 
systems that are chaotic in the classical limit. In the well-known kicked rotator a 
meaningful Zeno effect cannot occur due to the abrupt change of state vectors by 
the delta kicks. For fair evaluation of measurement-induced effects a new Hamiltonian 
with continuous time dependence has been considered instead. In this case,
it was shown that the shorter is the time interval of the measurement, the slower
is the decay of the survival probability. This clearly shows the quantum Zeno effect
occurs. We also examined this problem analytically and confirmed our conclusion.
Even though in the case of radiative or radioactive decay quantum anti-Zeno effect 
prevail \cite{Kofman00}, in the usual dynamical systems including chaotic ones 
quantum Zeno effect is more feasible.

We thanks Hyunchul Nha, Jaewan Kim, and Hai-Woong Lee for helpful discussion. 
This work is supported by Creative Research Initiatives of the Korean Ministry of 
Science and Technology.


\bibliographystyle{prsty}

\end{multicols} 

\end{document}